\documentstyle[aps,epsfig,float,twocolumn,prl]{revtex}
\newcommand{\be}{\begin{equation}}
\newcommand{\ee}{\end{equation}}

\begin{document}
\twocolumn[\hsize\textwidth\columnwidth\hsize\csname @twocolumnfalse\endcsname
\title{Lieb-Schultz-Mattis in Higher Dimensions}
\author{M. B. Hastings}
\address{
Center for Nonlinear Studies and Theoretical Division, Los Alamos National
Laboratory, Los Alamos, NM 87545, hastings@cnls.lanl.gov 
}
\date{July 28, 2003}
\maketitle
\begin{abstract}
A generalization of the Lieb-Schultz-Mattis theorem to higher dimensional
spin systems is shown.  The physical motivation for the result is that
such spin systems typically either have long-range order, in which
case there are gapless modes, or have only short-range correlations, in
which case there are topological excitations.
The result uses a set of
loop operators, analogous to those used in gauge theories, defined
in terms of the spin operators of the theory.  We also obtain
various cluster bounds on expectation values for gapped systems.  These
bounds are used, under the assumption of a gap, to rule out the first
case of long-range order, after which we show the existence of a topological
excitation.  
Compared to the ground state,
the topologically excited state has, up to a small error,
the same expectation values for all operators acting within any local
region, but it has a different momentum.
\vskip2mm
\end{abstract}
]
\section{Introduction}
Lieb, Schultz, and Mattis (LSM) proved in 1961 that a one-dimensional periodic
chain of length $L$, with half-integer spin per unit cell,
has an excitation gap bounded by ${\rm const.}/L$\cite{lsm}.  This
behavior contrasts with the possibility of a Haldane gap in the integer
spin case\cite{haldane}.

Despite several attempts\cite{try,try2}, this theorem has not been extended to 
higher dimensions.  The basic difficulty in obtaining a higher-dimensional 
version of this theorem was pointed out in two insightful papers by Misguich
and coworkers~\cite{ml}: 
if spin correlations are short-ranged, the ground
state wavefunction should be well described by a short-range 
resonating valence bond (RVB)
state\cite{rvb}.  The short-range RVB basis decomposes into
different topological sectors, depending upon the number of
dimers crossing a given line through the system.
This allows the construction of a low energy
excited state very similar to the twisted state of LSM\cite{bonesteel}.
Instead, if spin correlations are long-ranged,
such a state will not be low energy, but there will exist low energy spin
wave excitations.  In contrast to the one-dimensional case, there now exist
two distinct means of obtaining a low energy excitation, significantly 
complicating the proof of any such theorem. 

In the present paper, we show a higher-dimensional
version of the LSM theorem.
We consider a
$d$-dimensional system
of spin-$1/2$ spins, with
finite-range, $SU(2)$ invariant Hamiltonian $\cal{H}$, and
with an odd number of spins per unit cell on the lattice.  
Define the total number of unit cells in the lattice to be $V$.
Let $L$ be the number of unit cells in one particular direction, and let
$L$ be even; this direction will be referred to as the length.  
Therefore, $V$ is even (if $V$ were odd, there would be a trivial
spin degeneracy).
Let the system be periodic and translationally invariant in the
length direction.
Let $V/L^d$ be bounded by a constant $r$ (this constant $r$ is arbitrary,
and imposes some bound on the
behavior of the aspect ratio of the system).
Define $V/L$ to be the ``width" of the system, and let
this number be odd.
Then, we show that
if the ground state is unique,
the gap $\Delta E$ to the first excited state satisfies, 
\be
\label{bnd1}
\Delta E\leq c \log(L)/L,
\ee
where the constant $c$ depends on ${\cal H},d$, and where the result
holds for all $L$ greater than some minimum $L_0$, where $L_0$
depends on ${\cal H},d,r$\cite{hence}.

In this paper, we use the term gap to deal specifically with the
difference between the energy of the first excited state and the energy
of the ground state.  This includes two completely distinct physical
cases.  In the case of a one-dimensional system, a spin-$1/2$ Heisenberg
chain has a continuous spectrum of excitations above the ground state.
On the other hand, a Majumdar-Ghosh\cite{mg} chain 
has a doubly degenerate ground
state with a gap to the next excited state.  Weak perturbations
of the Majumdar-Ghosh Hamiltonian can break the exact degeneracy between the
two lowest states, leaving a system with a gap
from the ground state to the first excited state which is exponentially
small in the system size, and then a gap from the first excited state to
the next excited state which is non-vanishing even in the limit of large
system sizes.  We consider both of these cases as systems in which the gap
$\Delta E$ is vanishing in the limit of large system size.  Although
they are one-dimensional systems, these two cases closely match the two
possibilities mentioned above for higher-dimensional systems.  The first
case involves a system with a continuous spectrum as it has
algebraically decaying spin correlations.  In the second case, the
first excited state is very close to the twisted state of LSM.

The physical idea behind the proof of Eq.~(\ref{bnd1})
is closely related to the two possibilities considered above for the
absence of a gap.
In the event of long-range order, or algebraic long-range order, 
one expects that there is no gap.  Conversely,
if there is a gap, one expects that there is no long-range order.  This
is the first statement we prove: we assume that the system has a gap $\Delta E$
and show, in section III,
that connected expectation values decay exponentially in the
spacing between them.
Then, to prove Eq.~(\ref{bnd1}) we first assume that 
Eq.~(\ref{bnd1}) is violated, proceeding by contradiction.  
We assume the existence of such a gap violating Eq.~(\ref{bnd1}) in
sections IV, V, and VI.
We then use the existence of the gap $\Delta E$ and
the exponential decay of correlation functions to show an insensitivity of
the system to boundary conditions in section IV.  Then, this insensitivity is
used to construct a {\it low-energy}, twisted state in section V.  
The construction
of this twisted state will to some extent follow the topological 
attempt\cite{try2} at proving the LSM theorem in higher dimensions, with
some important differences outlined below.  
In section VI we will
show that the twisted state has a different momentum than the ground state.
It is here that the odd width of the system becomes essential.
Despite the different momentum compared to the ground state,
the twisted state has, up to a small error,
the same expectation values for all operators acting within any local
region.  Thus, we may refer to this state as a topologically excited state.
Finally, in an appendix, we briefly consider a version
of the result showing exponential decay of
correlation functions for systems governed by
certain Markov processes, rather than quantum systems.

Since we will constantly deal with operator equations of
motion, we introduce a set of ``loop operators" as
a basic technique.  
The loop operators, which will be suitably defined products of spin
operators, can be naturally interpreted as a product of gauge
fields around a loop\cite{ba}.  However, the use of these operators
avoids the uncontrolled approximations associated with the $U(1)$ and
$Z_2$ gauge theory techniques\cite{am}.  The introduction of these
operators is not necessary to the main development, but provides a
useful notation.

\section{Loop Algebra}
We define operators
$i^{\mu\nu}\equiv\frac{1}{2} \delta^{\mu\nu} + \sum_a
S_i^{a} \sigma_{a}^{\mu\nu}$, where $S_i^a$ are the spin operators
at site $i$
and $\sigma_a$ are the Pauli matrices, $a=x,y,z$.
Thus, $i^{\mu\nu}$ is the two-by-two matrix of spin operators
\be
\pmatrix{ \frac{1}{2}+S^z_i & S^x_i-i S^y_i \cr
S^x_i+i S^y_i & \frac{1}{2}- S^z_i \cr}
\ee

We consider operators of the form 
$i^{\mu\nu} j^{\nu\rho} k^{\rho\sigma} ... m^{\alpha\mu},$
which we refer to as loop operators,
where a summation over repeated indices is implied.
Later we will often suppress the indices $\mu,\nu$, writing $i$ rather
than $i^{\mu\nu}$ to save space.  Thus,
we will write the loop operator mentioned above in
the form ${\rm tr}(i j k ... m)$, 
where the trace $tr$ refers to a trace over
the Greek indices $\mu,\nu,...$.  Below we also use a
trace $Tr$; this trace $Tr$ refers to a trace of quantum operators, summing
over all states in the Hilbert space of the system.
Using the rule $i^{\mu\nu}i^{\rho\sigma}=\delta^{\nu\rho}i^{\mu\sigma}$, it
is always possible to reduce a given product of traces to a new 
product such that each site appears only once.
Then, an operator ${\rm tr}(i j k ... m)$ permutes the
spins around the sites $i,j...$ 

Given an operator $O(t)$, the operator obeys the equation of
motion $\partial_t O(t)=-i[O,{\cal H}]$.  
Consider, for example a term ${\rm tr}(ij)$ in ${\cal H}$.  We have
$[i,{\rm tr}(i j)]=ij-ji$.  As an illustration, let us give the full Greek
indices on this commutator: we have 
$[i^{\mu\nu},(i^{\alpha\beta}j^{\beta\alpha})]=
i^{\mu\alpha}j^{\alpha\nu}-
j^{\mu\alpha}i^{\alpha\nu}$, where summation over repeated indices
$\alpha,\beta$ is assumed.

Introduce coordinates $(x,y)$ to specify sites $i$, where
$x$ labels the unit cells along the direction of length and is 
defined up to integer multiples of $L$.  The coordinate $y$ labels the
unit cells along the other lattice directions, as well as labeling
the particular spin within the unit cell.  
Given two sites $i$ and $j$ on the lattice,
we define the distance between them, written $|i-j|$,
as the minimum number of moves by lattice vectors needed to
move from the unit cell containing $i$ to that containing $j$.  On
a square lattice, for example, this is the Manhattan distance.
Then, let $R$ denote the range of ${\cal H}$, the furthest distance between
two sites in any term in ${\cal H}$.  
If all distances in a product of loops are less than $L/2$, we can
define a winding number of the given product around the lattice
in the length direction.  If all distances in the product
remain less than $L/2-R$, then the dynamics ${\partial_t O}=-i[O,{\cal H}]$
does not connect sectors with different winding numbers.  We will make
use of the coordinates later by sometimes
writing loop operators ${\rm tr}(ij...)$ in the form
${\rm tr}((x_1,y_1)(x_2,y_2)...)$, where $i$ has coordinates $(x_1,y_1)$,
$j$ has coordinates $(x_2,y_2)$, and so on.

\section{Locality}
We consider ground state expectation values of operators $O_1,O_2...$, 
written $\langle O_1(t_1) O_2(t_2) ... \rangle$.  
The expectation values are not time ordered: the ordering of operators
is as written.  For a system with a unique
ground state and an energy gap $\Delta E$, on physical
grounds one expects that connected correlation functions,
defined as 
$\langle A(0) B(0) \rangle_c \equiv
\langle A(0) B(0) \rangle- \langle A(0) \rangle \langle B(0) \rangle$,
decay exponentially in distance
(without loss of generality, we will assume
$\langle A \rangle=\langle B \rangle=0$ through the rest of this section).
The proof of this locality bound will be done in this section.
We will do this in two steps: first, we consider commutators of
the form $[A(t),B(0)]$, where $A(0)$ and $B(0)$ are separated in space.
We bound the operator norm\cite{on} of the commutator for
sufficiently small $t$, and
thus bound its expectation value, in Eq.~(\ref{gdef}) below.  
The proof in this subsection will
just be sketched; a more rigorous derivation is due to Lieb and 
Robinson\cite{fgv}.  This result provides a bound on the velocity of the
system, as will be seen below.
Then, in the next subsection, from this bound on the
expectation value of the commutator and the existence of a gap, we
use a spectral representation of the commutator
to bound the connected correlation functions, thus obtaining the desired
locality bound on the expectation value, Eq.~(\ref{localbnd}).
Finally, we close the section by giving a similar locality bound for
operators separated in time.  

\subsection{Finite Velocity}
We define the distance between two operators $O_1,O_2$
to be $l$ if
the minimum distance between any pair of sites,
$i,j$, where $S_i$
appears in $O_1$ and $S_j$ appears in $O_2$, is $l$.
$A(0),B(0)$ are sums of products of spin or loop operators, which we
suppose to be distance $l$ apart.

We start with some notation.  The Hamiltonian, ${\cal H}$, can be
written as a sum of terms ${\cal H}=\sum_i {\cal H}_i$, such that
each ${\cal H}_i$ only contains spins operators on sites $j$ with
$|i-j|\leq R$.
Let $N_A$ denote the number of sites
appearing in $A(0)$, and $N_B$ denote the number of sites appearing
in $B(0)$.
Let $J$ denote the maximum, over sites $i$, of
$||{\cal H}_i||$.  

We now bound the operator norm of the commutator 
$[A(t),B(0)]$ for short times. 
On short time scales,
one expects that $A(t),B(0)$ are still separated in space, up to
small correction terms, as we now show.  
Consider first $||[A(t),{\cal H}_i]||$, and study the change
in this quantity as a function of time:
\begin{eqnarray}
\label{inh}
|\frac{||[A(t)-i{\rm d}t [A(t),{\cal H}],{\cal H}_i]||-||[A(t),{\cal H}_i]||}
{{\rm d}t}|
\\ 
\nonumber
=
|\frac{||[A(t),{\cal H}_i+i{\rm d}t [{\cal H}_i,{\cal H}]]||-||[A(t),{\cal H}_i]||}
{{\rm d}t}|\\ \nonumber \leq
\sum\limits_{|i-j|\leq 2R}
|\frac{||[A(t),{\cal H}_i+i{\rm d}t [{\cal H}_i,{\cal H}_j]]||-||[A(t),{\cal H}_i]||}
{{\rm d}t}| \\ \nonumber
=
\sum\limits_{|i-j|\leq 2R}
|\frac{||[A(t)-i{\rm d}t[A(t),{\cal H}_j],{\cal H}_i]||-||[A(t),{\cal H}_i]||}
{{\rm d}t}| \\ \nonumber
\leq
2 J\sum\limits_{|i-j|\leq 2R}
||[A(t),{\cal H}_j]||.
\end{eqnarray}
Here, we work to linear order in ${\rm d}t$.  While the operator 
$[A(t),{\cal H}_j]$ is differentiable, its operator norm need not be.
Thus, all equations here are correct when we take
the lim sup as ${\rm d}t\rightarrow 0$.

The first equality in Eq.~(\ref{inh}) is
obtained by moving the time derivative from $A(t)$ to ${\cal H}_i$ as follows:
for any operator $P$, to linear
order in ${\rm d}t$ we have $||P||=||P+i{\rm d}t [P,{\cal H}]||$.  
Set $P=[A(t)-i{\rm d}t[A(t),{\cal H}],{\cal H}_i]$.  Then, to linear
order in ${\rm d}t$, $||P||=||P+i{\rm d}t [P,{\cal H}]||=
||[A(t)-i{\rm d}t[A(t),{\cal H}],{\cal H}_i]+i{\rm d}t 
[[A(t),{\cal H}_i],{\cal H}]||=
||[A(t),{\cal H}_i+i{\rm d}t [{\cal H}_i,{\cal H}]]||$.

The inequality is obtained because $[{\cal H}_i,{\cal H}_j]=0$ for
$|i-j|>2R$.  The next equality is obtained by moving the time derivative
back to $A(t)$, using now the equality $||P||=||P-i{\rm d}t[P,{\cal H}_j]||$.
The final inequality results from the bound
$||{\cal H}_i||\leq J$.  

Now, let $S$ denote
the maximum number of sites $j$ within distance $R$ of any site $i$.
Eq.~(\ref{inh}) gives a set of differential
equations which bound the operator norm of various commutators; we
have also the initial conditions that $||[A(0),{\cal H}_j]||$ vanishes
for sites $j$ which are further than distance $R$ from any site in $A(0)$,
while $||[A(0),{\cal H}_j]||\leq 2 J ||A||$ for all other sites.  The
number of sites within distance $R$ of $A(0)$ is bounded by $N_A S$.

To bound $||[A(t),{\cal H}_j]||$, let us then consider the following set
of differential equations: for $t>0$, we take 
$\partial_t G_i=2 J\sum_{|i-j|\leq 2 R}
G_j$ and for $t<0$ we take
$\partial_t G_i=-2 J\sum_{|i-j|\leq 2 R}
G_j$, with initial conditions $G_j=0$ for sites $j$ which are
further than distance $R$ from any site in $A(0)$, and
$G_j=2 J ||A||$ for all other sites.  
Then, comparing these equations to Eq.~(\ref{inh}), we see that
$||[A(t),{\cal H}_i]||\leq G_i(t)$.
This set of linear equations for $G_i$ can be 
solved for any given lattice.  However, we are simply interested
in an upper bound on $G_i$.  Let us define $G^k$ to be the maximum of $G_i$ over
all sites $i$ which are at a distance greater than $(2k-1)R$ from all sites in
$A(0)$.  Then, we have $\partial_t G^k\leq 2 J S G^{k-1}$ for $k>0$, and
$\partial_t G^k\leq 2 J S G^k$ for $k=0$, with initial conditions
$G^0=2J||A||$ and $G^k=0$ for $k\geq 1$.
Thus, $G^0(t) \leq 2 J ||A|| e^{2 J S t}$, $G^1(t)\leq \int_0^{t} {\rm d}t'
2 J S G^0(t')\leq \int_0^{t} {\rm d}t' (2 J S) (2 J ||A||) e^{2 J S t} 
= 2 J ||A|| (2 J S t) e^{2 J S t}$, and 
$G^k(t)\leq 2 J ||A|| (2 J S t)^k e^{2 J S t}/k!$.  The
last set of inequalities follows inductively: $G^k(t) \leq
\int_0^t {\rm d}t' (2 J S) G^{k-1}(t) \leq 
\int_0^t {\rm d}t' (2 J S) (2 J ||A||)(2 J S t')^{k-1}/(k-1)!
e^{2 J S t}
=
2 J ||A|| (2 J S t)^k e^{2 J S t}/k!$. 
From these inequalities, 
we find, for a site $j$ which is at a distance greater than $(2 k-1) R$ 
from $A(0)$,
that 
$||[A(t),{\cal H}_j]||\leq 2 J ||A|| |2 J S t|^k e^{2 J S |t|}/k!$.

Finally, consider $\partial_t||[A(t),B(0)]||$.  Using a similar sequence of
inequalities to Eq.~(\ref{inh}), we find that
$\partial_t||[A(t),B(0)]||\leq 2 ||B|| \sum_j ||[A(t),{\cal H}_j]||$, where
the sum over $j$ extends over sites $j$ which are within distance $R$
of some sites in $B(0)$.  There are at most $N_B S$ such sites,
and each of them has $||[A(t),{\cal H}_j]||\leq G^{k-1}(t)$, where
$k=l/2R$.  Here, we take $k$ to be the ceiling of $l/2R$, the smallest integer
greater than or equal to $l/2R$; we obtain this value of $k$ since each
such site is at least a distance 
$l-R$ from all sites in $A(0)$, and
so we need $l-R>[2(k-1)-1]R$.
Then, 
$||[A(t),B(0)]||\leq 2 N_B ||A|| ||B|| |2 J S t|^{l/2R} 
e^{2 J S |t|}/(l/2R)!$.

Define $f(t)\equiv\langle [A(t),B(0)] \rangle$.
Since $f(t)\leq ||[A(t),B(0)]||$,
\begin{eqnarray}
\label{gdef}
f(t)
\leq \frac{2 N_B ||A|| ||B|| |2 J S t|^{l/2R} e^{2 J S |t|}}{(l/2R)!}
\\ \nonumber \equiv N_B ||A|| ||B|| g(t,l).
\end{eqnarray}
For $t=c_1 l$, the large $l$ behavior of $g(t,l)\sim
\exp[(l/R) (2 J S c_1 R + 1/2 + (1/2){\log}(4 J S c_1 R))]$.  If we choose
a sufficiently small $c_1$, then $g(c_1 l,l)$
decays exponentially in $l$ for large $l$.  Numerically, we find that the
zero of
$2 J S c_1 R + 1/2 + (1/2){\log}(4 J S c_1 R))]$, is at
$c_1\approx 0.139232/(2 J S R)$.  Any $c_1$ smaller than this value
(for example,
$c_1=0.1/(2JSR)$ will work) will cause $g(c_1 l,l)$ to be exponentially
decaying for large $l$.
The velocity at which correlations spread in the system is of order
$c_1^{-1}$.

\subsection{Spectral Decomposition}
Now, we use a spectral decomposition of $f(t)$ to relate $f(t)$ to
the desired correlation function, $\langle A(0) B(0) \rangle$.
Without loss of generality, let us set the ground state energy, $E_0$,
to $0$.
The spectral decomposition of $f(t)$ gives 
\be
\label{spectral}
f(t)=\sum_{i} (e^{-iE_i t}A_{0i}B_{i0}
-e^{iE_i t}B_{0i}A_{i0}),
\ee
where $A_{i0}$ is the matrix element of
operator $A$ between the ground state $0$ and the eigenstate $i$, 
with
energy $E_i\geq \Delta E$ 
above the ground state energy, and similarly for the other
$A_{0i},B_{i0},B_{0i}$.  There are no terms in Eq.~(\ref{spectral})
involving $A_{00},B_{00}$ since we have assumed $\langle A \rangle=
\langle B \rangle = 0$.

Let us define a function $f^+(t)=\sum_{i} e^{-E_i t}A_{0i}B_{i0}$, which
thus contains only the negative frequency (positive energy) terms in $f(t)$.
The significance of $f^+(t)$ is that $f^+(t)=\langle A(t) B(0)\rangle$,
so that the positive energy part of $f(t)$ contains the desired
correlation function in it.  In this subsection, we combine the
bound (\ref{gdef}) on $f(t)$ with the existence of a gap to bound
$f^+(0)$.

Define $\tilde f(t)=f(t) e^{-t^2 \Delta E^2/(2 q)}$, with $q$ to
be chosen later.  
We have two bounds on $\tilde f(t)$.
First, we have the bound (\ref{gdef}) on $f(t)$ which gives
us the bound $\tilde f(t)\leq N_B ||A|| ||B|| g(t,l)$.  We also have
have 
\be
\label{tldbnd}
\tilde f(t)\leq 2 ||A|| ||B|| e^{-t^2 \Delta E^2/(2 q)}
.
\ee
We will
use the first of these bounds for times $|t|<c_1 l$, and the second for long
times $|t|>c_1 l$.
Finally, we define ${\tilde f}^+(t)$ to contain only the
negative frequency terms in $\tilde f(t)$.

Now, the desired expectation value $\langle A(0) B(0) \rangle=f^+(0)$.
To bound $f^+(0)$, we
first bound ${\tilde f}^+(0)$, and then bound
${\tilde f(0)}^+-f^+(0)$.
To bound ${\tilde f}^+(0)$, we use the bounds on
$\tilde f$ and an integral representation of the positive energy 
part\cite{mdef1,contour}:
\begin{eqnarray}
\label{intrep}
|{\tilde f}^+(0)|=\frac{1}{2\pi}|\int_{-\infty}^{\infty} {\rm d}t\,
\tilde f(t)/(-i t+\epsilon)|\\ 
\nonumber
=\frac{1}{2\pi}\Bigl(
|\int_{|t|<c_1 l} \tilde f(t)/(-i t+\epsilon)|+
|\int_{|t|>c_1 l} \tilde f(t)/(-i t+\epsilon)|\Bigr) \\ \nonumber
\leq \frac{1}{2\pi}
||A|| ||B|| \Bigl(2 N_B g(c_1 l,l)+ 2 \frac{\sqrt{2\pi q}}{\Delta E c_1 l}
e^{-c_1^2 l^2 \Delta E^2/(2q)}\Bigr).
\end{eqnarray}
In Eq.~(\ref{intrep}), to
bound the integral over $|t|<c_1 l$, we
used $|\tilde f(t)|\leq |f(t)| \leq N_B ||A|| ||B|| g(t,l)
\leq N_B ||A|| ||B|| [|t|/(c_1 l)] g(c_1 l,l)$.  To derive this inequality we
have assumed that $l>0$ so that taking the ceiling of $l/2R$ above gives
a $k\geq 1$.
Then,
\begin{eqnarray}
|\int_{|t|<c_1 l} {\rm d}t \tilde f(t)/(-it+\epsilon)|\leq  \\ \nonumber
N_B ||A|| ||B|| \int_{|t|<c_1 l} {\rm d}t g(c_1 l,l)/(c_1 l)=
\\ \nonumber
2 N_B ||A|| ||B||
g(c_1 l,l).
\end{eqnarray}
To bound the integral over $|t|>c_1 l$ in Eq.~(\ref{intrep})
we have used Eq.~(\ref{tldbnd}) to show
$|\int_{|t|>c_1 l} \tilde f(t)/(-it+\epsilon)|\leq
2||A|| ||B|| \int_{-\infty}^{\infty}{\rm d}t
\exp[-t^2\Delta E^2/(2 q)]/(c_1 l)\leq
2 ||A|| ||B|| \frac{\sqrt{2 \pi q}}{\Delta E c_1 l}
\exp[-c_1^2 l^2 \Delta E^2/(2 q)]$.

To bound
$|{\tilde f(0)}^+-f(0)^+|$, we start with
the definition of $\tilde f$.  
Expressed as a convolution in Fourier space\cite{mdef2} this is:
\be
\label{conv}
\tilde f(\omega')=
(\sqrt{2\pi q}/\Delta E)
\int d{\omega}
f(\omega) e^{-q (\omega-\omega')^2/(2\Delta E^2)}
.
\ee

Now is where the existence of an energy gap becomes essential.
For motivation,
let us first pictorially (see Fig.~1) describe how the gap enables us to bound
$|{\tilde f^+(0)}-f^+(0)|$ and then present it more mathematically.
By definition $\tilde f(0)=f(0)$; this follows in Fourier space from
$\int_{-\infty}^{\infty} {\rm d}\omega f(\omega)=
\int_{-\infty}^{\infty} {\rm d}\omega \tilde f(\omega)$.
The convolution (\ref{conv}) means that a given Fourier component in
$f$ which is, for example, negative frequency, will produce both positive
and negative frequency Fourier components in $\tilde f$.  So, consider
a $\delta$-function spike in $f(\omega)$, produced by an intermediate
state $i$ with energy $E_i=-\omega>0$.  This produces a Gaussian in
$\tilde f(\omega)$, as shown.  The integral over all $\omega$ of the
Gaussian is the same as the integral of the $\delta$-function; however,
the shaded portion of the curve has $\omega>0$.  Since
$\tilde f^+(0)=(2\pi)^{-1} \int_{-\infty}^{0} {\rm d}\omega \tilde f(\omega)$
and
$f^+(0)=(2\pi)^{-1} \int_{-\infty}^{0} {\rm d}\omega f(\omega)$,
we find a difference between $\tilde f^+(0)$ and $f^+(0)$ equal to
the integral of the shaded portion of the curve.  
At $\omega=0$ the height of the Gaussian is reduced
by a factor $e^{-q \omega^2/(2\Delta E^2)}$.
However, since $E_i\geq \Delta E$, this factor is bounded by
$e^{-q/2}$.

Now, let us do the calculation more directly:
$\tilde f^+(0)=(2\pi)^{-1} \int_{-\infty}^{\infty} {\rm d}\omega
f(\omega) \int_{-\infty}^{0} {\rm d}\omega' 
(\sqrt{2 \pi q}/\Delta E)
\exp[-q(\omega-\omega')^2/(2\Delta E^2)]$, while
$f^+(0)=(2\pi)^{-1}\int_{-\infty}^{0} {\rm d}\omega f(\omega)$.
Then 
\be
\label{gapb}
\tilde f^+(0)-f^+(0)=(2\pi)^{-1} \int_{-\infty}^{\infty} {\rm d}\omega
f(\omega) [\Theta_q(-\omega)-\Theta(-\omega)].
\ee
Here
$\Theta(\omega)$ is a step function: $\Theta(\omega)=1$ for $\omega>0$ and 
$\Theta(\omega)=0$ for $\omega<0$.  We have defined
\be
\label{thqd}
\Theta_q(\omega)=
\int_{0}^{\infty} {\rm d}\omega' 
(\sqrt{2 \pi q}/\Delta E)
\exp[-q(\omega-\omega')^2/(2\Delta E^2)].
\ee
Since
the system has a gap, the integral in Eq.~(\ref{gapb}) vanishes for
$|\omega|<\Delta E$.  However, for $|\omega|\geq \Delta E$, we
have $|\Theta_q(-\omega)-\Theta(-\omega)| \leq
e^{-q/2}$.  Thus, Eq.~(\ref{gapb}) is bounded by
\be
\label{rb}
(2\pi)^{-1} e^{-q/2} \int_{-\infty}^{\infty} {\rm d}\omega  |f(\omega)| \leq
2 ||A|| ||B|| e^{-q/2}.
\ee

Thus, combining Eqs.~(\ref{intrep},\ref{rb}),
$|f^+(0)|\leq |{\tilde f}^+(0)-f^+(0)|+|{\tilde f}^{+}(0)|
\leq 
\frac{1}{2\pi} ||A|| ||B|| (2 N_B g(c_1 l,l) +
2\frac{\sqrt{2\pi q}}{\Delta E c_1 l}
e^{-c_1^2 l^2 \Delta E^2/(2q)})
+2 ||A|| ||B|| e^{-q/2}$.
We finally choose $q=c_1 l \Delta E$ to get
\begin{eqnarray}
\label{localbnd}
|\langle A(0) B(0) \rangle_c |\leq \frac{1}{2\pi} 
2 N_B ||A|| ||B|| g(c_1 l,l) + \\ \nonumber
2 (1+\frac{1}{\sqrt{2 \pi c_1 l \Delta E}}) ||A|| ||B||
e^{-c_1 l\Delta E/2},
\end{eqnarray}
giving the desired bound.  The
first term in Eq.~(\ref{localbnd}) decays as
$\exp[-{\cal O}(l/R)]$,
while 
the second term decays as $\exp[-c_1 l\Delta E/2]=
\exp\{-{\cal O}[\Delta E l/(JSR)]\}$; here, by ${\cal O}(l/R)$, we mean
some quantity of order $l/R$.

In what follows in the next three
sections, the first term in Eq.~(\ref{localbnd}) will be negligible: we
will be considering operators separated by a distance $l$ which is of
order $L$, so that the first term in Eq.~(\ref{localbnd}) will lead to
only exponentially small (in $L$) contributions to the correlation functions.
The second term will be more important: since we will consider gaps
$\Delta E\propto \log(L)/L$, the second term will lead to terms which
are suppressed only by powers of $L$ when considering correlation functions
of operators separated by a distance of order $L$.

\subsection{Operators at Different Times}
It is possible to extend the result Eq.~(\ref{localbnd}) 
to correlation functions
$\langle A(0) B(i\tau) \rangle$, with $\tau$ real and $\tau>0$.
Then, in Eq.~(\ref{intrep}), we must evaluate
$|\tilde f^+(-i\tau)|$, so that the denominator $(-it+\epsilon)$ is replaced
by $-it+\tau$.  In this case, we are still able to find just as tight
a bound on $|\tilde f^+(-i\tau)|$ as we previously found for
$|\tilde f^+(0)|$: $|\tilde f^+(-i\tau)|\leq
\frac{1}{2\pi}
||A|| ||B|| \Bigl(2 N_B g(c_1 l,l)+ 2 \frac{\sqrt{2\pi q}}{\Delta E c_1 l}
e^{-c_1^2 l^2 \Delta E^2/(2q)}\Bigr)$.

Of course,
for $\tau\geq q/\Delta E$ there is the trivial bound
$|\langle A(0) B(i\tau) \rangle_c|\leq ||A|| ||B|| e^{-\tau \Delta E}\leq
||A|| ||B|| e^{-q}$. 
For $|\tau|\leq q/\Delta E$, we claim that
$|{\tilde f}^{+}(-i\tau)-\exp[+\tau^2\Delta E^2/(2 q)]
f^+(-i\tau)|\leq 2 ||A|| ||B|| e^{-q/2}$.
To show this, 
$|{\tilde f}^{+}(-i\tau)-
\exp[+\tau^2\Delta E^2/(2 q)]
f^+(-i\tau)|=$
\begin{eqnarray}
\label{iqz}
(2\pi)^{-1}\int_{-\infty}^{\infty} {\rm d}\omega f(\omega)
\Bigl(
\int_{-\infty}^{0}
{\rm d}\omega' (\sqrt{2\pi q}/\Delta E)
\\ \nonumber
\exp[\omega' \tau] \exp[-q(\omega-\omega')^2/(2\Delta E^2)]
- \\ \nonumber
\exp[+\tau^2\Delta E^2/(2 q)]
\Theta(-\omega)\exp[\omega \tau]\Bigr).
\end{eqnarray}
The portion of the integral with $\omega<0$ is equal to
\begin{eqnarray}
\label{iqz2}
(2\pi)^{-1}\int_{-\infty}^{-\Delta E} {\rm d}\omega f(\omega)
\int_0^{\infty} {\rm d}\omega'
(\sqrt{2\pi q}/\Delta E) \\ \nonumber
\exp[\omega' \tau] \exp[-q(\omega-\omega')^2/(2\Delta E^2)]
,
\end{eqnarray}
where we have 
used the gap $\Delta E$ and the relation
$\int_{-\infty}^{\infty} 
{\rm d}\omega'(\sqrt{2\pi q}/\Delta E)
\exp[\omega' \tau] \exp[-q(\omega-\omega')^2/(2\Delta E^2)]\\=
\exp[\omega \tau] \exp[+\tau^2\Delta E^2/(2 q)]$.
Then, for $\tau\leq q/\Delta E$, the integral (\ref{iqz2}) with
$\omega<0$ is
bounded in absolute value by $(2\pi)^{-1} \int_{-\infty}^{0} {\rm d}\omega
|f(\omega)|
e^{-q/2}$.  We can similarly bound the portion of the integral with
$\omega>0$, giving the desired result.

With the given $q=c_1l\Delta E$
the above bounds show that for $\tau\leq c_1 l$,
$|\langle A(0) B(i\tau) \rangle_c|\leq$
\begin{eqnarray}
\label{tsep}
e^{-\tau^2\Delta E/(2 c_1 l)}
[\frac{1}{2\pi} 
2 N_B ||A|| ||B|| g(c_1 l,l) + \\ \nonumber
2 (1+\frac{1}{\sqrt{2 \pi c_1 l \Delta E}}) ||A|| ||B||
e^{-c_1 \Delta E l/2}].
\end{eqnarray}

\section{Twisted Boundary Conditions}
In this section we derive some results on the sensitivity to boundary
conditions, as a step towards the the main result, Eq.~(\ref{bnd1}).  To derive
a contradiction later, we will assume throughout this and the
next two sections that there is a gap
$\Delta E$ that violates Eq.~(\ref{bnd1}), with an appropriately
chosen $c$.  In the first subsection, we review the twist of boundary
conditions and the topological attempt at proving the LSM theorem.
In the second subsection, we show the specific results on the
sensitivity to boundary conditions.

\subsection{Topological Argument}
Here we will define a new twisted Hamiltonian, making use of
the coordinates, $(x,y)$, introduced previously for lattice sites $i$.
To define the new twisted
Hamiltonian, ${\cal H}_{\theta,\theta'}$, replace
all loop operators
${\rm tr}((x_1,y_1)(x_2,y_2)(x_3,y_3) ... )$ in ${\cal H}$ 
with ${\rm tr}((x_1,y_1)R(x_1,x_2)(x_2,y_2)R(x_2,x_3)(x_3,y_3)...)$.
Here, the twist operator $R^{\mu\nu}(x_1,x_2)\equiv
\exp[\pm i\frac{\theta}{2}\sigma_z^{\mu\nu}]$ if the shortest lattice path
between $x_1,x_2$ crosses from $x=0$ to $x=1$, where the sign is positive if the
path crosses in the direction of increasing $x$ and negative if it crosses
in the opposite direction.  Here, $R^{\mu\nu}(x_1,x_2)$ is a
two-by-two matrix of numbers, rather than of operators,
\be
R^{\mu\nu}=\pmatrix{\exp[\pm i\frac{\theta}{2}] & 0 \cr
0 & \exp[\mp i\frac{\theta}{2}] \cr}
\ee

Alternately, if the shortest lattice path
between $x_1,x_2$ crosses from $L/2$ to $L/2+1$,
$R^{\mu\nu}(x_1,x_2)\equiv \exp[\pm i\frac{\theta'}{2}\sigma_z^{\mu\nu}]$.
Otherwise,
$R(x_1,x_2)=\delta^{\mu\nu}$.   In Fig.~(2), we show the coordinate system
using $x,y$ and show where the two boundary condition twists are inserted.

Let us see what the effect of this twist is in terms of spin operators.
Consider two sites, $i,j$.  Suppose the Hamiltonan
${\cal H}_{0,0}$ has a term such as
${\rm tr}(ij)=2 (S^z_i S^z_j + S^x_i S^x_j + S^y_i S^y_j) + 1/2$.
Then, let us suppose $i$ has $x=0$ while $j$ has $x=1$.  Then,
${\cal H}_{\theta,\theta'}$ has a term ${\rm tr}(i R(0,1) j R(1,0))$.  In terms
of spin operators, this is equal to 
$2 [S^z_i S^z_j + \cos(\theta)(S^x_i S^x_j + S^y_i S^y_j)
+\sin(\theta)(S^x_i S^y_j-S^y_i S^x_j)] + 1/2$.  In the untwisted Hamiltonian,
we coupled the dot product of the two spin vectors, $\vec S_i,\vec S_j$; in
the twisted Hamiltonian, we couple them after rotating one by an angle
$\theta$ about the $z$-axis.
A good discussion of twists can be found in \cite{ml}.

We have considered two different twist angles, $\theta,\theta'$.
The spectrum of ${\cal H}_{\theta,\theta'}$ 
depends only on the combination $\theta+\theta'$.  Further,
from any given eigenfunction $\psi(\theta,\theta')$ of 
${\cal H}_{\theta,\theta'}$, one can find an eigenfunction 
$\psi(\theta-\delta\theta,\theta'+\delta\theta)$ of 
${\cal H}_{\theta-\delta\theta,\theta'+\delta\theta}$ by
$\psi(\theta-\delta\theta,\theta'+\delta\theta)=
\prod_j e^{i\delta\theta S^z_j}\psi(\theta,\theta')$, where the product extends
over all sites $j=(x,y)$ with $0<x\leq L/2$.

Given that the spectrum depends only on the combination
$\theta+\theta'$, the reader may wonder what the reason is for introducing
two twist angles, rather than just one angle.  In fact, the second angle is
a useful trick, introduced for the following reason:
we have previously shown that the existence of a gap causes
correlation functions to decay exponentially in the separation of
the two operators.  However, physically, one expects that the existence of a gap
will also imply some insensitivity of the system to boundary conditions,
enabling us to bound, for example, the second derivative of the
ground state energy with respect to $\theta$.  What we will do in
the next subsection is show this insensitivity by using the fact that the
spectrum depends only on $\theta+\theta'$ to convert the second derivative
($\partial_{\theta}^2$)
of the ground state
energy into a mixed partial derivative
($\partial_{\theta}\partial_{\theta'}$) of the ground state energy, 
and by then evaluating that mixed
partial derivative as a correlation function, using the
exponential decay of correlation functions.  This will be stated more
precisely at the start of the next subsection; we mention it here for
motivation.

The eigenvalues are of $\cal{H}_{\theta,\theta'}$ are invariant under 
$\theta+\theta'\rightarrow\theta+\theta'+2\pi$,
while the wavefunctions are invariant under 
$\theta+\theta'\rightarrow\theta+\theta'+4\pi$\cite{try2,ml}.
To motivate the results in this section, we recall the basic idea of
the topological attempt\cite{try2} at proving the LSM.  The idea is
that if there is a gap at $\theta+\theta'=0$, and {\it if the gap remains open}
for all $\theta+\theta'$, then under an adiabatic change in the angle $\theta$
with $\theta'$ fixed at zero,
the ground state at $\theta=0$ evolves into the ground state at
$\theta=2\pi$.  At $\theta=2\pi$, the Hamiltonian is returned to the
original Hamiltonian, but, for a system of odd width, the ground state
expectation
value of the translation operator changes sign, as will be discussed in
more detail below.  This leads to a contradiction: from the ground state
with given expectation value of the translation operator, we construct
another ground state with the opposite expectation value.  The
requirement that the topological attempt requires the gap to remain open
for all $\theta$ was pointed out in \cite{ml}.

What the topological argument actually succeeds in showing is that the
gap must close at some value of $\theta$.  However,
in order to use this argument to obtain any bound on the magnitude of the gap 
at $\theta=0$,
we would have to show that a sufficiently large gap at $\theta=0$
would prevent the gap from closing for all $\theta$; that
would then lead to a contradiction, enabling us to bound the gap
at $\theta=0$.  What we
will see is that we can partially show this:
for sufficiently large $c$ in Eq.~(\ref{bnd1}),
we can show to second order in $\theta$ (or indeed, to any
finite order)
a bound on the change in ground state energy with respect to $\theta$.
However, we will be unable to show that the gap remains open for
all $\theta$ because to bound the change in ground state
energy for higher orders in $\theta$
requires progressively increasing the constant $c$ in Eq.~(\ref{bnd1}),
and it is not possible to show the result to all orders.  Thus, the
topological attempt will ultimately fail, and we will give a physical
example of how this can happen.  In the next section (V), we will give
a successful argument.

\subsection{Boundary Condition Sensitivity}
We now show an insensitivity of the ground state energy, $E_0(\theta,\theta')$,
to second order\cite{diffable} in the twist angle, $\theta+\theta'$.  
At $\theta=\theta'=0$, $\partial_{\theta} E_0(\theta,\theta')=
\langle \partial_{\theta} {\cal H}_{\theta,\theta'}\rangle=0$.
Indeed, taking any {\it odd} number of derivatives of $E_0(\theta,\theta')$
leads to a vanishing quantity\cite{rotg}.
To second order in $\theta,\theta'$, we write a power series:
$E_0(\theta,\theta')=E_0(0,0)
+a \theta^2/2 + a \theta'^2/2 + b \theta \theta'$, where 
$a=\partial_{\theta}^2 E_0=\partial_{\theta'}^2 E_0$ and 
$b=\partial_{\theta\theta'} E_0$.  
We will show that, for any given negative power of $L$,
we can find a constant $c$ such if Eq.~(\ref{bnd1}) is violated for that $c$,
then $a$ is bounded by an ${\cal H}$-dependent constant times the
given negative power of $L$.
We do this by calculating $b$ as a correlation function, and
then showing that $b=a$.

Recall linear perturbation theory:
suppose a Hamiltonian ${\cal H}$ is changed by some $\delta {\cal H}$.
For a non-degenerate state, $|\psi\rangle$, with eigenvalue $E$, 
the change $|\delta \psi\rangle$ 
in $|\psi\rangle$ is given to linear order
in $\delta {\cal H}$ by
$|\delta \psi\rangle=(E-{\cal H})^{-1}\delta{\cal H}|\psi\rangle$.
Since the ground state is the lowest energy state, all other states
have energies greater than it.  Thus, we can write the change in
the ground state to linear
order as $|\delta\psi_0\rangle=-\sum_{a\neq 0} \int_0^{\infty} 
{\rm d}\tau
e^{(E_0-E_a)\tau}
|\psi_a\rangle
\langle\psi_a|\delta{\cal H}|
\psi_0\rangle
=-\int_0^{\infty} 
{\rm d}\tau
\delta{\cal H}(i\tau)|\psi_0\rangle $, where
$|\psi_0\rangle$ is the ground state wavefunction, $|\psi_a\rangle$
are a complete set of intermediate states, 
and where 
$\delta{\cal H}(i \tau)=\exp[-{\cal H}\tau]\delta{\cal H}
\exp[{\cal H}\tau]$
is the change in the Hamiltonian operator, taken at imaginary time $i\tau$.
Here we have set $E_0(0,0)=0$ without loss of generality.

Specializing to the case of $\delta {\cal H}=\partial_{\theta}
{\cal H}_{\theta,\theta'}$
and writing the change in $\psi_0$
in terms of the $\theta,\theta'$-dependent
ground state density matrix 
$\rho^0(\theta,\theta')\equiv |\psi_0(\theta,\theta')
\rangle\langle\psi_0(\theta,\theta')|$ we have:
\begin{eqnarray}
\label{gt}
\partial_{\theta}\rho^0
=-\int_0^{\infty} {\rm d}\tau\,\partial_{\theta}
{\cal H}_{\theta,\theta'}(i\tau)\rho^0
-\int_{-\infty}^{0} {\rm d}\tau\,\rho^0
\partial_{\theta}{\cal H}_{\theta,\theta'}(i\tau), \\ \nonumber
\partial_{\theta'}\rho^0
=-\int_0^{\infty} {\rm d}\tau\,\partial_{\theta'}
{\cal H}_{\theta,\theta'}(i\tau)\rho^0
-\int_{-\infty}^{0} {\rm d}\tau\,\rho^0
\partial_{\theta'}{\cal H}_{\theta,\theta'}(i\tau).
\end{eqnarray}
Note that since $\langle \delta {\cal H} \rangle$ vanishes in this case,
we do not need to worry about matrix elements of $\delta {\cal H}$ from
the ground state to the ground state.

Now, we can use the change in the density matrix to compute $b$
by $b={\rm Tr}[(\partial_{\theta'}{\cal H}_{\theta,\theta'})
(\partial_{\theta}\rho^0)]$.
So,
\begin{eqnarray}
\label{bcorr}
b=-\int_{0}^{\infty} {\rm d}\tau \,
\Bigl(\langle 
\partial_{\theta} {\cal H}_{\theta,\theta'}(0) 
\partial_{\theta'} {\cal H}_{\theta,\theta'}(i \tau) \rangle
\\ \nonumber
-\langle\partial_{\theta'} {\cal H}_{\theta,\theta'}(-i\tau) 
\partial_{\theta} {\cal H}_{\theta,\theta'}(0) 
\rangle\Bigr),
\end{eqnarray}
where the derivatives are evaluated at $\theta=\theta'=0$.
The derivative $\partial_{\theta}{\cal H}_i$ is non-vanishing only
for sites $i$ which are within distance $R$ of $x=0$; there are at most
$SV/L$ such sites.  For each $i$, 
$||\partial_{\theta}{\cal H}_i||\leq JS$, so
$||\partial_{\theta}{\cal H}_{\theta,\theta'}||\leq JS^2V/L$.
We use two bounds for the given correlation functions in Eq.~(\ref{bcorr}).
First, each correlation function is bounded by
$(JS^2V/L)^2 e^{-\tau \Delta E}$.  Second, we can use Eq.~(\ref{tsep}) to
bound each correlation function by
\be
\label{bcf}
2 S (JS^2V/L)^2 
(1+1/\sqrt{\pi c_1 L \Delta E}) \exp[-c_1 \Delta E (L/2)/2],
\ee
where we neglect
the term in $g(c_1 l,l)$ in Eq.~(\ref{tsep}) 
as it leads to a correction which is
exponentially decaying in
$L$, not in $c_1 \Delta E L$, and thus is negligible in what follows.  Also, 
we have used $l=L/2$, ignoring the slight error, that in fact $l\geq L/2-R$.
Finally, we have used $N_B \leq S$ in Eq.~(\ref{bcf}).

Using these two bounds on the correlation function, 
we arrive at
\begin{eqnarray}
|b| \leq 2 (JS^2 V/L)^2 \int\limits_0^{\infty} {\rm d}\tau \, {\rm min}\{
\exp[-\tau \Delta E],x\},
\end{eqnarray}
where $x=2S(1+1/\sqrt{\pi c_1 L \Delta E}) \exp[-c_1 \Delta E (L/2)/2]$.
Thus, $|b|\leq 2 (JS^2 V/L)^2 (x/\Delta E)(1+\log(x))$.
The number of sites, $V$, is bounded by $r L^d$, while
for $\Delta E$ greater than
$c{\rm ln}(L)/L$, 
$x$ is
bounded by a $c$-dependent
negative power of $L$.  Therefore, we can bound
$|b|$ by any desired negative power of $L$
by choosing $c$ sufficiently large.

However, $E_0(\theta,-\theta)=E_0(0,0)$, so
$b=a$.  Thus, we have also bounded $|a|$ by
the same negative power of $L$.  
Therefore, at $\theta=\theta'=0$, we find that 
$|\partial_{\theta}^2 E_0(\theta,\theta')|$ is
bounded by a negative power of $L$.
This shows some insensitivity of the
ground state energy to boundary conditions.
This realizes the physical idea\cite{ml} that a spin liquid
state is defined by the lack of response to a twist in boundary conditions to 
second order in $\theta$\cite{sl}.  

At fourth order in $\theta$, 
we must evaluate a correlation function of {\it four} operators, each of
order $JS^2 V/L$; to bound these correlation functions requires
a larger $c$.  Each higher order in $\theta,\theta'$ requires
an even larger $c$, so that it is not possible to
bound the change in ground state energy for arbitrary $\theta+\theta'$.
Therefore, the topological attempt\cite{try2} to establish the LSM
result fails.  Indeed,
a gap at $\theta+\theta'=0$ must close for 
$\theta+\theta'\neq 0$\cite{ml}.

It is worth giving a specific physical example of this possibility,
as the topological argument does show that the gap must close
for some $\theta+\theta'$.
In many
physical examples of spin liquids, the closing of the gap arises because
a state which is at some very low energy, of order $JL^{-1}$ or less,
 above the ground state at
$\theta+\theta'=0$ crosses the ground state energy at a finite $\theta+\theta'$.
For example, if the Majumdar-Ghosh Hamiltonian is slightly perturbed, there
is a state at an exponentially small energy above the ground state which
crosses the ground state at $\theta+\theta'=\pi$.

However, it is also possible for a state which is at some energy $JL^{0}$ to
cross the ground state: consider a system with two competing phases,
one of which is a spin liquid phase while the other is a spin ordered phase
with a spiral order.  The spiral order is chosen so that the spin ordered
phase can be frustrated at $\theta+\theta'=0$, and the spin liquid is
the ground state there.  At some $\theta+\theta'\neq 0$, however, the
spiral phase can take over as the ground state.  This taking over as
the ground state can happen either via a level crossing (if the two states
have different symmetry, for example, or if the spin ordered phase has
a non-vanishing net spin), or via an avoided crossing.
This provides a specific
example of a system in which a state or phase which is at an energy of order 
$JL^0$ at $\theta+\theta'=0$ becomes the ground state at some non-vanishing
$\theta+\theta'$.

The solution to this problem is simple: it is not necessary to show that
there is a gap for all twist angles.  Instead, we start
with the ground state at vanishing twist and continuously evolve this
state, obtaining a state for any twist angle which is an approximate
eigenstate of the twisted Hamiltonian, not necessarily the ground state.
This approximate eigenstate will be explicitly
constructed in the next section, while in the section after that
we demonstrate that at a twist of $2\pi$ the expectation value of
the translation operator has changed sign in the new state compared to
the ground state.  Thus, this gives a new low energy state, different
from the ground state.  

\section{Twisting the Ground State}
\subsection{Constructing the Twisted State}
Let $\rho(\theta,0)$ be a $\theta$-dependent density
matrix that we construct below.
Divide the system into two overlapping
halves: half (1) contains sites with
$x=3L/4-R,3L/4-R+1,...,L-1,0,1,...L/4+R$, while half (2) contains sites
with $x=L/4-R,L/4-R+1,...,3L/4+R$.  That is, half (1) contains all sites
from $x=3L/4-R$ up to $x=L-1$, as well as all sites from $x=0$ up
to $L/4+R$, while half (2) contains all sites from $x=L/4-R$ up to $x=3L/4+R$.
The halves are shown as shaded regions in Fig.~(3).  

The reason we choose
two overlapping halves is that we will be considering density
matrices which involve only sites within a given half.  These matrices
will be defined by tracing over sites outside the given half.  Then, to
evaluate the expectation value of the
energy of the system, we will be able to evaluate the expectation value 
as a sum of operators which lie completely within one or the other half.  That 
is, by making the two halves overlap, we will deal with the question of
the ``seam" where the two halves join.  This is mentioned here as motivation
and will be done in more detail below.

Define $\rho_1(\theta,0)={\rm Tr}_{2}[\rho(\theta,0)]$, where
${\rm Tr}_2$ denotes a trace over all sites {\it not in half (1)}, and define
$\rho_2(\theta,0)={\rm Tr}_{1}[\rho(\theta,0)]$,
the trace over sites not in half (2).
Similarly, define 
$\rho_1^0(\theta,\theta')={\rm Tr}_{2}[\rho^0(\theta,\theta')]$, and
$\rho_2^0(\theta,\theta')={\rm Tr}_{1}[\rho^0(\theta,\theta')]$.
We will assume throughout this section that there is
a gap violating Eq.~(\ref{bnd1}).  Then, for
sufficiently large $c$, we will construct $\rho_1$ such that
\begin{eqnarray}
\label{dp}
\rho_{1}(\theta,0)-E_1(\theta)=
\Bigl( \prod_{j} e^{i\theta S^z_j} \Bigr)
\rho_{1}^0(0,0) \Bigl(\prod_{j} e^{-i\theta S^z_{j}} 
\Bigr)\\ \nonumber
=\rho_{1}^0(\theta,-\theta),
\end{eqnarray}
where the
products extend over all sites $j=(x,y)$ with $0<x\leq L/4$
and where $E_1(\theta)$ is an error term such that the
trace norm\cite{tn} $|E_1(\theta)|$ is 
bounded by a constant times
a negative power of $L$ for all $0\leq \theta \leq 2\pi$.  The particular
negative power of $L$ can be determined by choosing the constant $c$ in 
Eq.~(\ref{bnd1}).  
As a useful terminology, we will refer to a quantity
as ``small" if, for any desired negative power of $L$, we can
find sufficiently large $c$ or sufficiently large
$q$ (introduced below), such that the given quantity is
bounded by a constant times the given negative power of $L$ for all $L$.
Thus, we wish $|E_1(\theta)|$ to be small.  Note that, given this definition
of small, if a small quantity is multiplied by any fixed power of $L$, the
result is a small quantity.  Sometimes, we will indicate that a quantity
is made small by choosing $c$ or by choosing $q$, to specify which of
the two needs to be made sufficiently large.

In differential form, we require
\be
\label{ddp}
\partial_{\theta}\rho_1(\theta,0)=
\sum_j i[S^z_j,\rho_1^0(\theta,-\theta)]+e_1(\theta),
\ee
where $e_1=\partial_{\theta}E_1$.  We will
show that the upper\cite{upper} derivative
${\cal D}_{\theta}|E_1(\theta)|$ is small,
from which 
Eq.~(\ref{dp}) will follow.
We will also require $\rho_2(\theta,0)=\rho_2^0(0,0)$, up to a similarly
bounded error term $E_2(\theta)$, and 
$\partial_{\theta}\rho_1(\theta,0)=e_2(\theta)$, with a similarly bounded
${\cal D}_{\theta}|E_2(\theta)|$.

The physical motivation behind Eq.~(\ref{dp}) is to construct
a state for the Hamiltonian ${\cal H}_{\theta,0}$ that has an
energy close to $E_0(0,0)$.  The twist 
$\theta$ is along a line that lies completely within half (1) while
$\theta'$ is along a line that lies completely within half (2).  
Within half (1), the Hamiltonians ${\cal H}_{\theta,0}$ and 
${\cal H}_{\theta,-\theta}$ are equal, so we construct a density matrix
such that
{\it within half (1)} the given density matrix
is close to the ground state density matrix of ${\cal H}_{\theta,-\theta}$.
Then, the expectation of any operator $O$ which lies completely within half
(1) for the density matrix $\rho(\theta,0)$ will be within $||O|| |E_1(\theta)|$
of the expectation value of that operator for the density matrix 
$\rho(\theta,-\theta)$.
On the other hand, within half (2), the Hamiltonians ${\cal H}_{\theta,0}$
and ${\cal H}_{0,0}$ are equal, so we also require that within half (2)
the density matrix be close to the ground state density matrix of
${\cal H}_{0,0}$.

Then, the expectation
value of the energy in the state defined by $\rho(\theta,0)$ is equal to
${\rm Tr}[\rho(\theta,0){\cal H}_{\theta,0}]$.
Once we have shown that both Eq.~(\ref{dp}) and the bound on
$|E_2(\theta)|$ are satisfied, it will follow that this
expectation value
will be within an amount $||{\cal H}_{\theta,0}|| {\rm max}(|E_1|,|E_2|)$ 
of $E_0(0,0)$, since the Hamiltonian ${\cal H}$ can be written as a sum of
operators which are entirely within half (1) or entirely within half (2)
(it was for this reason that the halves were chosen to overlap).  
Therefore,
since $||{\cal H}||$ is bounded by $V J<r L^d J$,
if we pick $c$ in Eq.~({\ref{bnd1}) sufficiently large,
we will find that
${\rm Tr}[\rho(\theta,0){\cal H}_{\theta,0}]-E_0(0,0)$ will also be small
at $\theta=2\pi$; this follows from the statement that a small quantity
multiplied by a fixed power of $L$ is also small.

Our claim, which we show in this section,
is that Eq.~(\ref{dp}) is satisfied by a
$\rho(\theta,0)$ defined as follows for $0\leq \theta\leq 2\pi$.  We pick
\be
\label{eqm2}
\partial_{\theta}\rho(\theta,0)=
-\int\limits_0^{c_1 L}
{\rm d}\tau \, [A^+(i\tau)-A^-(-i\tau),\rho(\theta,0)],
\ee
where we define 
\begin{eqnarray}
\label{adef}
A^+(i\tau)=(2\pi)^{-1}\exp[-(\tau\Delta E)^2/(2q\log(L))] \times \\ \nonumber
\int_{-\infty}^{\infty} {\rm d}t\,
\partial_{\theta}{\cal H}_{\theta,0}(t)
\exp[-(t\Delta E)^2/(2q\log(L))]/
(-it+\tau),
\end{eqnarray}
with
$q$ to be chosen later, and $A^-(-i\tau)=(A^+(i\tau))^{\dagger}$.  
The time evolution of the operator
$\partial_{\theta}{\cal H}_{\theta,\theta'}(t)$
is defined using the Hamiltonian ${\cal H}_{\theta,-\theta}$, while
the $\tau$ dependence of $A^{+,-}$ is defined via Eq.~(\ref{adef}).

To give some insight into the definition of $A^+,A^-$, we note that if
$q$ were to be infinite, then they would project onto positive and
negative energy parts of $\partial_{\theta}{\cal H}$ at times $\pm i\tau$,
respectively.
That is, for $q=\infty$, we have 
$A^+(i\tau)=(2\pi)^{-1} \int_{-\infty}^{\infty} {\rm d}t 
\partial_{\theta}{\cal H}_{\theta,\theta'}(t)/(-it+\tau)$. 
Let the matrix elements of the
operator $\partial_{\theta}{\cal H}$ in a basis of eigenstates
of ${\cal H}$ be written $(\partial_{\theta}{\cal H})_{ab}$ where the
states have energies $E_a,E_b$.  Let the states have energy difference
$-\omega=E_a-E_b$.  Then, doing the integral over $t$ we find that
$A^+(i\tau)$ has a matrix element between states $a,b$ equal to
$\exp[\omega\tau]
(\partial_{\theta}{\cal H})_{ab}$ for $-\omega>0$ and equal to zero
for $-\omega<0$.  Similarly, $A^-(-i\tau)$ has a matrix element equal
to $\exp[-\omega\tau]
(\partial_{\theta}{\cal H})_{ab}$ for $-\omega<0$ and equal to zero
for $-\omega>0$.  Then, for any given time $\tau$, the integrand of
Eq.~(\ref{eqm2}) would be the same as that of Eq.~(\ref{gt}) for
$\theta=0$, since in that case the only non-vanishing terms in
Eq.~(\ref{eqm2}) are $-A^+(i\tau)\rho(0,0)-\rho(0,0)A^-(-i\tau)$.

What we will do later is to take a finite $q$ instead.  Physically, this
means that rather than taking an adiabatic change in $\theta$
which keeps us in the ground state, we instead ``pass through" the level
crossing when the gap closes at some $\theta\neq 0$, 
going from the ground state to some low energy
excited state.

Eq.~(\ref{eqm2}) gives the change in $\rho$ equal to the commutator of
$\rho$ with an anti-Hermitian operator, and hence generates an infinitesimal
unitary transformation of $\rho$.  Thus, $\rho$ continues to be a density
matrix which projects onto a single state, defined to be $\psi(\theta,0)$.

As a first step,
we wish to show that for $\theta=\theta'=0$ we can find a $c$
such that
$\partial_{\theta}\rho_1^0(\theta,0)-\sum_j i[S^z_j,\rho_1^0(0,0)]$ is small.
We have 
$\partial_{\theta}\rho_1^0(\theta,0)-\sum_j i[S^z_j,\rho_1^0(0,0)]=$ 
\be
\label{untwist}
\partial_{\theta}\rho_1(\theta,0)-
\partial_{\theta}\rho_1^0(\theta,-\theta)=
{\rm Tr}_2[\partial_{\theta'}\rho^0(0,\theta')],
\ee
where all derivatives are evaluated at $\theta=\theta'=0$.
To bound the right-hand side of Eq.~(\ref{untwist}), 
consider the trace of this term
with any operator $O$ with $||O||=1$.  
This operator must be within half (1), so, using
Eq.~(\ref{gt}) to compute the derivative of $\rho^0$ with respect to
$\theta'$,  we obtain the expectation value ${\rm Tr}[O\partial_{\theta'}
\rho^0(0,\theta')]=$
\begin{eqnarray}
\label{ds}
\partial_{\theta'}\langle O \rangle= 
-\int\limits_{0}^{\infty} {\rm d}\tau
\, \langle O \partial_{\theta'}{\cal H}_{\theta,\theta'}(i\tau)+
\partial_{\theta'}{\cal H}_{\theta,\theta'}(-i\tau) O\rangle.
\end{eqnarray}
However,
following the arguments from the previous section and the
locality bounds, we can find a $c$
such that Eq.~(\ref{ds}) is small.  In this case, the distance between
$O$ and $\partial_{\theta'}{\cal H}$ is at least $L/4-2R$, since
$\partial_{\theta'}{\cal H}$ includes terms with $x$ down to $L/2-R$, while
$O$ is in half (1) so includes $x$ up to $L/4+R$.
Note that $\langle \partial_{\theta'}{\cal H}_{\theta,\theta'}\rangle=0$
at $\theta=\theta'=0$.  Since we have bounded the trace of the right-hand side
of Eq.~(\ref{untwist}) with all operators $O$ with $||O||=1$, 
we have bounded the trace norm of the right-hand side.

\subsection{Bound on Error Terms}
We now show that we can find a $c$ such that the definition (\ref{eqm2})
satisfies Eq.~(\ref{ddp}) in general.
We wish to compute
$\partial_{\theta}\rho_1(\theta,0)-\sum_j i[S^z_j,\rho_1(\theta,-\theta)]
=e_a(\theta)+e_b(\theta)+e_c(\theta)$.  Here we define
\begin{eqnarray}
\label{ea}
e_a\equiv
-\int\limits_0^{c_1 L}{\rm d}\tau \, 
{\rm Tr}_2
[[A^+(i\tau)-A^-(-i\tau),\rho(\theta,0)]- \\ \nonumber
[A^+(i\tau)-A^-(-i\tau),\rho^0(\theta,-\theta)]].
\end{eqnarray}
\begin{eqnarray}
\label{eb}
e_b\equiv -\int\limits_0^{c_1 L}{\rm Tr}_2[
[A^+(i\tau)-A^-(-i\tau),\rho^0(\theta,-\theta)]] \\ \nonumber
-\partial_{\theta}\rho_1^0(\theta,\theta').
\end{eqnarray}
\begin{eqnarray}
\label{ec}
e_c\equiv
\partial_{\theta}\rho_1^0(\theta,\theta')-
\sum_j i[S^z_j,\rho_1^0(\theta,-\theta)].
\end{eqnarray}
In Eq.~(\ref{eb}), the derivative of $\rho_1^0$ is evaluated at 
$\theta=-\theta'$. 
We now consider each of these terms $e_a,e_b,e_c$ in
turn.  

First, consider Eq.~(\ref{ea}).  In the definition of $A^{+,-}$ as an integral 
over $t$, the integral over times $|t|>c_1 (L/2-R)$ has an operator norm
bounded by $||\partial_{\theta}{\cal H}|| \int_{|t|>c_1 (L/2-R)} 
\exp[-(t\Delta E)^2/(2q\log(L)]$.  Thus, for any fixed $q$ (to be chosen later)
we can find a $c$
such that this integral over times $|t|>c_1 (L/2-R)$ has small
operator norm, and thus when commuted with $\rho(\theta,0)$ gives
a term with small trace norm.

Eq.~(\ref{ea}) involves an integral
of $\partial_{\theta}{\cal H}(t)$
over time $t$ in the definition of $A^{+,-}$; 
we have shown that the contributions with times 
$|t|>c_1 (L/2-R)$ may be neglected.  Then, considering only contributions
with $|t|\leq c_1 (L/2-R)$, we claim that, up to an error in
the operator norm of order $\exp[-{\cal O}(L)]$,
$\partial_{\theta}{\cal H}(t)$ can be written as an operator
involving only terms not in half (2).  That is, that
$||\partial_{\theta}{\cal H}(t)-{\rm Tr}_2[\partial_{\theta}{\cal H}(t)]||$
is exponentially small in $L$.  To show this, define $U_{12}$ to
be the set of all
sites $j$ which lie in both half (1) and half (2); there
are at most $2SV/L$ such sites.  These
sites are shown in the solid 
regions in Fig.~(3).  Define operators $O(t=0)=O'(t=0)
=\partial_{\theta}{\cal H}$, and define the time evolution
of $O,O'$ by $\partial_t O=-i[O,{\cal H}]$, while 
$\partial_t O'=-i\sum_{i\not \in U_{12}} [O,{\cal H}_i]$, i.e., the time
evolution of $O'$ includes only the sum over sites $i$ which are either in
half (1) or in half (2), but {\it not} in both halves.
Then, using
the arguments leading up to Eq.~(\ref{gdef}), we can show that
for $i\in U_{12}$, the operator norm
$||[O(t),{\cal H}_i]||$ is bounded by $2J||O||g(c_1(L/2-R),L/2-R)$, which
is of order $\exp[-{\cal O}(L/2-R)]$ for the given range of times $t$.
Then, using the difference in the evolution equations for $O,O'$,
we can bound $||O(t)-O'(t)||\leq \sum_{i\in U_{12}}
\int_0^t {\rm d}t' ||[O,{\cal H}_i]||\leq 2t(2SV/L)J||O||g(c_1(L/2-R),L/2-R)$.
This quantity is also of order 
$\exp[-{\cal O}(L/2-R)]$ for the given range of times $t$.  Finally,
we use the fact that ${\rm Tr}_2[O']=O'$ to get the
desired result.

From the above two paragraphs, it follows that up to
small error in the trace norm,
$e_a(\theta)=
-\int\limits_0^{c_1 L}{\rm d}\tau \, 
[{\rm Tr}_2[A^+(i\tau)-A^-(-i\tau)],{\rm Tr}_2[\rho(\theta,0)-
\rho^0(\theta,-\theta)]]$.
Then, this is
equal to the commutator of $E_1(\theta)$ with
an anti-Hermitian operator.  It generates an infinitesimal unitary rotation of
$E_1(\theta)$ and therefore
does not lead to any change in $|E_1(\theta)|$.

Next, consider Eq.~(\ref{eb}).  First consider the terms in the commutator
involving $A^{+,-}$ acting on the left side of $\rho^o$.
As above, the operator $\partial_{\theta}{\cal H}$
can be written in a basis of eigenstates of ${\cal H}$ as
$(\partial_{\theta}{\cal H})_{ab}$, where the states have energies $E_a,E_b$.
In $\partial_{\theta}{\cal H}\rho^0(\theta,-\theta)$
the only non-vanishing terms involve
states with energy difference $-\omega=E_a-E_b\geq \Delta E$.
Consider a matrix element $(\partial_{\theta}{\cal H})_{ab}$
with given $\omega$.  This leads to a matrix element of $A^{-}(-i\tau)$
equal to 
$(\partial_{\theta}{\cal H})_{ab}$ times
\begin{eqnarray}
\label{wconvm}
\frac{\sqrt{2\pi q\log(L)}}{\Delta E}
\exp[-(\tau \Delta E)^2/(2q\log(L)] 
\times \\ \nonumber
\int_{-\infty}^{\infty}
\frac{{\rm d}\omega'}{2\pi} \, \Theta(\omega') \exp[-\omega'\tau] 
\exp[-q\log(L)(\omega-\omega')^2/(2\Delta E^2)],
\end{eqnarray}
where we have
converted the time integral to an integral in Fourier space.
Since $-\omega\geq\Delta E$,
Eq.~(\ref{wconvm}) can be made small
by choosing $q$ sufficiently large.
Thus, the trace norm of $A^-(-i\tau)\rho^0(\theta,-\theta)$
is small, for all
$\tau\geq 0$.  Similarly, for $A^{+}$, we find that we get a matrix element
equal to
$(\partial_{\theta}{\cal H})_{ab}$ times
\begin{eqnarray}
\label{wconvp}
\frac{\sqrt{2\pi q\log(L)}}{\Delta E}
\exp[-(\tau \Delta E)^2/(2q\log(L)] 
\times \\ \nonumber
\int_{-\infty}^{\infty}
\frac{{\rm d}\omega'}{2\pi} \, \Theta(-\omega') \exp[\omega'\tau] 
\exp[-q\log(L)(\omega-\omega')^2/(2\Delta E^2)].
\end{eqnarray}
By choosing $q$ sufficiently large,
the integral (\ref{wconvp}) can be made
equal to $\exp[\omega\tau]$, up to small
error.
Thus, the given matrix element
can be made
equal to $(\partial_{\theta}{\cal H})_{ab}$ times $\exp[\omega \tau]$,
up to small error.
Therefore, the trace norm of
$-\int\limits_0^{c_1 L}{\rm d}\tau\,{\rm Tr}_2\{
[A^+(i\tau)-\partial_{\theta}{\cal H}(i\tau)]\rho^0(\theta,-\theta)\}$
is small.  These statements amount to saying that,
with small error in the operator norm, 
$A^+(i\tau)$ indeed is equal to the positive energy part
of $\partial_{\theta}{\cal H}(i\tau)$, while
$A^-$ is equal to the negative energy part.

Now consider $A^{+,-}$ acting to
the right side of $\rho^0(\theta,-\theta)$, so that we consider
$\rho^0(\theta,-\theta)\partial_{\theta}{\cal H}$.  
In that case, the only non-vanishing
terms in $(\partial_{\theta}{\cal H})_{ab}$ involve 
$-\omega=E_a-E_b\leq -\Delta E$.
Repeating the argument above, we find that the trace norm of
$\rho^0(\theta,-\theta) A^+(i\tau)$ is small, as is the trace norm of
$\int\limits_0^{c_1 L}{\rm d}\tau\,{\rm Tr}_2\{
\rho^0(\theta,-\theta)
[A^-(-i\tau)-\partial_{\theta}{\cal H}(-i\tau)]
\}$.

Therefore, up to small error, Eq.~(\ref{eb}) is equal to
$-\int\limits_0^{c_1 L}{\rm d}\tau\,{\rm Tr}_2[
\partial_{\theta}{\cal H}(i\tau)\rho^0(\theta,-\theta)
+\rho^0(\theta,-\theta) \partial_{\theta}{\cal H}(-i\tau)]
-
\partial_{\theta}\rho_1^0(\theta,\theta')$, which equals
$-\int\limits_0^{c_1 L}{\rm d}\tau\,{\rm Tr}_2[
\partial_{\theta}{\cal H}(i\tau)\rho^0(\theta,-\theta)
+\rho^0(\theta,-\theta) \partial_{\theta}{\cal H}(-i\tau)]
+
\int\limits_0^{\infty}{\rm d}\tau\,{\rm Tr}_2[
\partial_{\theta}{\cal H}(i\tau)\rho^0(\theta,-\theta)
+\rho^0(\theta,-\theta) \partial_{\theta}{\cal H}(-i\tau)]$.
This difference is equal to an integral over $\tau\geq c_1 L$.  For
sufficiently big $c$, the trace norm of
this integral can be bounded by any desired
negative power of $L$.  Thus, $e_b$ has small trace norm.

Finally, consider Eq.~(\ref{ec}).  This is equal to
\be
\label{unt2}
\partial_{\theta}\rho^0_1(\theta,\theta')-\partial_{\theta}
\rho^0_1(\theta,-\theta)={\rm Tr}_2[\partial_{\theta'}\rho^0(\theta,\theta')],
\ee
where the derivatives are evaluated at $\theta'=-\theta$.
The trace norm of the right-hand side of Eq.~(\ref{unt2}) can be bounded
by a negative power of $L$ using the same arguments near Eq.~(\ref{ds}),
by considering
an operator $O$ that is entirely within half (1).  The only difference
to the arguments near Eq.~(\ref{ds}) is that we compute the derivatives
and expectation values at $\theta=-\theta'$, rather than at $\theta=\theta'=0$.

Thus, using Eqs.~(\ref{ea},\ref{eb},\ref{ec}) and picking
sufficiently large $q$ we find that
${\cal D}_{\theta}|E_1(\theta)|$ is small
for sufficiently large $c$.
A similar sequence of arguments permits one to bound ${\cal D}_{\theta}
|E_2(\theta)|$.
In the next section, we consider the expectation value of the translation
operator on $\rho(2\pi,0)$.

\section{Translation Operator}
Consider the operator ${\rm tr}((1,y)(2,y) 
... (L,y))$, which translates the sites with given $y$.  The translation
operator $T$
which translates the entire system by one unit cell
is the product of these loop operators over all $y$ (there are an odd number
of such loop operators).  
The ground state of ${\cal H}_{0,0}$ is an eigenstate of $T$.
If the ground state is non-degenerate, then it has eigenvalue
$\pm 1$; without loss of generality we will assume in this section
that is has eigenvalue $+1$.

In this section we will show that the expectation value of $T$ for
$\rho(2\pi,0)$ is opposite to that for $\rho(0,0)$, up to
small error.  We note
that if ${\cal H}_{\theta,0}$
were to have a gap for all $0\leq\theta\leq 2\pi$, then the results in this
section would provide the last step in the topological argument discussed
above.  Instead, the results in this section will complete the argument started
in the previous section:
$\rho(2\pi,0)$ gives us a density matrix
such that
${\rm Tr}[\rho(2\pi,0){\cal H}_{\theta,0}]-E_0(0,0)$ is small,
but which, up to small error,
has the opposite expectation value for $T$.  Since the difference
in the expectation of ${\cal H}$ is small, we can find a $c$ such that
the difference in expectation value decays faster than $1/L$,
and then we can find an $L_0$ such that
for $L>L_0$ the state $\psi(2\pi,0)$ has an energy expectation value which
is {\it less than} $c{\rm ln}(L)/L$ above the ground state.  However,
since the expectation value of $T$ is opposite for $\rho(2\pi,0)$ compared
to $\rho(0,0)$, up to small error, this state has an overlap on the
ground state which is small.
Thus, we will show in this section a contradiction under the assumption
that the system had a gap $\Delta E$ which violated Eq.~(\ref{bnd1})
and under the assumption that the system was translation symmetric, so
that the ground state was an eigenstate of $T$.

We first define a twisted translation operator, $T_{\theta,\theta'}\equiv
\prod_y {\rm tr}(\exp[i\frac{\theta}{2}\sigma_z]
(1,y)(2,y)...(L/2,y)
\exp[i\frac{\theta'}{2}\sigma_z]
(L/2+1,y)...(L,y))$.
Then, $T_{\theta,\theta'}$ is a unitary operator
and a symmetry of ${\cal H}_{\theta,\theta'}$.
Finally, given that $T\psi_0(0,0)=\psi_0(0,0)$, we have
$T_{\theta,-\theta}\psi_0(\theta,-\theta)=\psi_0(\theta,-\theta)$ for
all $\theta$.

We will then show that
$\partial_{\theta}|T_{\theta,0}\psi(\theta,0)-\psi(\theta,0)|$
is small for all $0\leq \theta \leq 2 \pi$.
It will then follow that, up to a small error,
${\rm Tr}[\rho(0,0)T_{0,0}]={\rm Tr}[\rho(2\pi,0)T_{2\pi,0}]=
-{\rm Tr}[\rho(2\pi,0)T_{0,0}]=1$,
thus showing that the twisted state indeed has the opposite expectation
value for $T$.
Here we have used the fact that
for systems of {\it odd} width, $T_{2\pi,0}=-T_{0,0}$.

Consider the derivative
$\partial_{\theta}(T_{\theta,0}\psi(\theta,0)-\psi(\theta,0))=$
\begin{eqnarray}
i\sum\limits_{y} S^z_{(1,y)} T_{\theta,0}\psi(\theta,0)-\\ \nonumber
\int\limits_0^{c_1 L}
{\rm d}\tau \, [
T_{\theta,0}A^+(i\tau)
-A^+(i\tau) \\ \nonumber
-T_{\theta,0}A^-(-i\tau)
+A^-(-i\tau)
]\psi(\theta,0).
\end{eqnarray}
This is equal to
\begin{eqnarray}
\label{dot}
\{
i\sum\limits_{y} S^z_{(1,y)} 
-\int\limits_0^{c_1 L}
{\rm d}\tau \, [
T_{\theta,0}A^+(i\tau)T_{\theta,0}^{-1}-A^+(i\tau) \\ \nonumber
-T_{\theta,0}A^-(-i\tau)T_{\theta,0}^{-1}+A^-(-i\tau)]\}
\psi(\theta,0)+ \\ \nonumber
\{
i\sum\limits_{y} S^z_{(1,y)} -
\int\limits_0^{c_1 L} {\rm d}\tau \,
[T_{\theta,0}A^+(i\tau)T_{\theta,0}^{-1}
\\ \nonumber
-T_{\theta,0}A^-(-i\tau)T_{\theta,0}^{-1}]\}
\left\{T_{\theta,0}\psi(\theta,0)-\psi(\theta,0)\right\}
\end{eqnarray}

The last term of Eq.~(\ref{dot}) is equal to an anti-Hermitian operator
acting on $T_{\theta,0}\psi(\theta,0)-\psi(\theta,0)$, and thus does
not change the norm of this state.  Thus, we need to bound the norm of the
first
term.  This term is equal to an anti-Hermitian
operator, that we define to be $O_1$, acting on $\psi$.  The norm square
of this term is equal to ${\rm Tr}[O_1^{\dagger} O_1 \rho(\theta,0)]$.  As
shown in the previous section, up to small 
error in the operator norm, $A^{+,-}(i\tau)$ can be written entirely
as operators in half (1).  Therefore, 
$O_1$ can be written entirely as an operator in half (1); that is,
the operator norm $||O_1-{\rm Tr}_2[O_1]||$ is 
small.  Thus, the
norm square is, up to small error, ${\rm Tr}[O_1^{\dagger} O_1
\rho_1(\theta,0]$,
which, again up to small error, is equal to
${\rm Tr}[O_1^{\dagger} O_1 \rho_1^0(\theta,-\theta)]$, since
$|\rho_1(\theta,0)-\rho_1^0(\theta,-\theta)|$ is small.  

We claim, however, that this last expectation value is small.
To show this, consider
$\partial_{\theta}(T_{\theta,-\theta}\psi_0(\theta,-\theta)-
\psi_0(\theta,-\theta))$.  This is equal to zero.  However, this
derivative can be written as an operator $O$ acting on
$\psi_0$, with $O=i\sum_y (S^z_{(1,y)}-S^z_{(L/2+1,y)})-
\int_0^{\infty} {\rm d}\tau\, (
T_{\theta,-\theta}
\partial_{\theta}{\cal H}_{\theta,\theta'}(i\tau)
T_{\theta,-\theta}^{-1}-
\partial_{\theta}{\cal H}_{\theta,\theta'}(i\tau)
-
T_{\theta,-\theta}
\partial_{\theta'}{\cal H}_{\theta,\theta'}(i\tau)
T_{\theta,-\theta}^{-1}+
\partial_{\theta'}{\cal H}_{\theta,\theta'}(i\tau)
)$, where
the derivatives are taken at $\theta=-\theta'$.
Since $O\psi_0=0$, it follows that
${\rm Tr}[O^{\dagger}O\rho^0(\theta,-\theta)]=0$.
However, up to small error,
$O=O_1+O_2$, with $O_1$ the operator considered above and $O_2$ defined
to be a similar operator acting only in half (2).  Then,
$2{\rm Tr}[O_1^{\dagger} O_1 \rho^0(\theta,-\theta)]+
2{\rm Tr}[O_1^{\dagger} O_2 \rho^0(\theta,-\theta)]=0$.  However,
using the locality bounds, the second term can be made small for large enough
$c$\cite{h12}, 
and thus the first term,
${\rm Tr}[O_1^{\dagger} O_1 \rho_1^0(\theta,-\theta)]$, is small.  Therefore, we
have shown the desired result.

\section{Discussion}
The main result is Eq.~(\ref{bnd1}), obtaining a bound on the energy
gap for spin models in arbitrary dimensions.  In order to obtain this result,
we have introduced a set of loop operators, and
proven a bound on
connected correlation functions.  This bound on correlation functions
did not rely on the system being a spin system; rather, it is valid for
any Hamiltonian such that the ${\cal H}_i$ have bounded operator norm,
and such that the interaction is finite range.
Below, we generalize this bound on
correlation functions to certain
other systems as well.

We note that
for the case of higher spin representations of $SU(2)$, Eq.~(\ref{bnd1}) follows
automatically from the result for spin-$1/2$, so long
as the total spin within all unit cells is half-odd: the higher spins can
be written as various combinations of spin-$1/2$ spins, and if the
total spin in the unit cell is half-odd then there will result
an odd number
of spin-$1/2$ spins in each unit cell.  Suppose, for example, a unit
cell contains one spin-$1$ spin and one spin-$1/2$ spin, giving a total spin of
$3/2$ which is half-odd.  Then, the spin-1 can be written
as two spin-$1/2$ spins.
Let these two spins be called $S_1,S_2$ and let the Hamiltonian
include only terms symmetric under interchange of $S_1,S_2$.  
This new Hamiltonian has three spin-$1/2$ spins in each unit cell, and
hence falls within the class of Hamiltonians considered above.
Then, there are two different sectors of the Hilbert space with no terms in the
Hamiltonian coupling these two sectors:
one sector in which $S_1,S_2$ form a spin-$0$, and
one in which they form a spin-$1$.  By adding a term coupling $S_1$ to
$S_2$ to the Hamiltonian with a sufficiently large, negative (ferromagnetic)
coefficient, 
we can ensure that the ground and first excited states lie in the sector
in which $S_1,S_2$ have total spin-$1$.  Then, the existence of a low-lying
state satisfying Eq.~(\ref{bnd1}) for the new system with only 
spin-$1/2$ implies
the existence of such a low-lying state for the original 
system with both spin-$1$ and spin-$1/2$.
It would also be interesting to generalize these results
to other groups $SU(N)$,
as well as to consider the case of even $V/L$.

We finish with two conjectures.  First, we conjecture that the same
Eq.~(\ref{bnd1}) holds for systems with an even width, so long as
the width $V/L$ is of order $L^{d-1}$ and so long as $d>1$.  For $d=1$,
this result is of course not true, as Haldane gap behavior is possible.

Second, consider the {\it thermal}
expectation value of $T$ at an inverse temperature $\beta$, defined
by $\langle T \rangle_{\beta}\equiv
{\rm Tr}[\exp(-\beta {\cal H}) T]/{\rm Tr}[\exp(-\beta {\cal H})]$.
We conjecture that there is a constant $c$, depending on ${\cal H},d$ such
that for $\beta>c\log(L)/L$ the given thermal expectation value 
$\langle T \rangle_{\beta}$ vanishes in the limit $L\rightarrow \infty$
for systems of odd width.  We base this conjecture on the following 
physical observations: for ferromagnetic systems, there are spin
wave excitations, with dispersion relation $E\propto k^2$.  It may be
shown that
the presence of these excitations causes $\langle T \rangle_{\beta}$
to vanish for $\beta$ of order $L$ as $L \rightarrow \infty$.  
For antiferromagnetic systems,
the translation symmetry is broken by the antiferromagnetic ordering
(in fact, for these systems, the true ground state has translation symmetry
and is a superposition of different broken symmetry states, but there
are low-lying
states with different expectation values of $T$ so that for $\beta$ of
order $L$, the expectation value $\langle T \rangle_{\beta}$ vanishes).
Finally, for spin liquid systems, there is a low-lying excited state
with the opposite expectation value of $T$ compared to the ground state,
as we have found above.  We leave a proof of both of these conjectures for
future work.

\appendix\section{Markov Processes and Locality}
Consider a system with a probability
$p_i$ of being in state $i$ and a transition matrix $T_{ij}$ so that
the equation of motion is $\partial_t p_i=\sum_j T_{ij} p_j$.  For the
total probability to be conserved, we have $\sum_i T_{ij}=0$, which
guarantees that $T_{ij}$ has at least one zero eigenvalue.  Let us assume,
further, that all eigenvalues of $T_{ij}$ are real.  This includes
all systems for which the stationary state (given by the zero eigenvector
of $T_{ij}$) obeys detailed balance.
A typical example of such a process would be the Monte Carlo dynamics of
a statistical mechanics system.  We will first derive a suitable
generalization of the locality result (\ref{localbnd}) to
systems governed by such a Markov process, and then discuss the
implication for statistical mechanics systems.

Let us assume that the spectrum of $T$ is such that there is only
one zero eigenvalue, with right eigenvector $p_i^0$, and that
all other eigenvalues $\lambda$ are negative with $\lambda\leq -\Delta$,
for some $\Delta>0$.  Assume $p_i^0$ is normalized by $\sum_i p_i^0=1$.

Then, introduce various quantities to be measured, $A, B,...$, so that
the expectation value of $A$ is given by $\langle A \rangle=
\sum_i A_i p_i^0$.  We can write this slightly differently by
introducing for each such quantity a diagonal matrix given by
$\hat A_{ii}=A_i$ for all $i$ and $\hat A_{ij}=0$ for
$i\neq j$.  Further, introduce an additional vector
$I_i$, such that $I_i=1$ for all $i$.  This vector $I_i$ is a left
eigenvector of $T$ with zero eigenvalue, since $\sum_i T_{ij}=0$ as
mentioned above.  Then,
$\langle A \rangle=I_i A_{ij} p_j^0$.

We can now consider expectation values of quantities at different
times: $\langle  A(t) B(0) \rangle \equiv I^{\dagger}
\hat A \exp[T t] \hat B p^0=
I^{\dagger} \exp[-T t] \hat A \exp[T t] \hat B p^0=I^{\dagger}
\hat A(t) \hat B(0) p^0$.  In these equations, $I^{\dagger}$ denotes the
transpose of the vector $I$ ($I$ is real, so no complex conjugation
is necessary), and we have left off all the indices on
vectors $I,p$ and matrices $A, B, \exp[\pm T t]$: the product is
evaluated following the usual rules of matrix multiplication.
In the sequence of equalities above,
the first equality defines the time evolution of the system, the
second equality follows since $I \exp[-T t]=I$, and the last equality
follows since we define $\hat A(t)$ by the equation of motion:
$\partial_t \hat A(t)=[A(t),T]$.
It is then possible to extend this definition to operators
separated by an {\it imaginary} time separation: $\langle A(it) B(0) \rangle$.

Now, consider a typical physical example: an Ising system, governed by
Monte Carlo spin flip dynamics, with $A(0),B(0)$ representing the
value of two different spins which are separated in space.  In such
a case (as well as in many others), it is possible to obtain a bound
similar to Eq.~(\ref{gdef}).  Assume that the matrix $T$ can be
written as a sum of matrices $T_i$, with finite interaction range $R$ and 
with a bound $||{T}_i||\leq J$, where $i$ is a site index.  
Define $f(t)\equiv\langle [A(it),B(0)] \rangle$.
Since $f(t)\leq ||[A(it),B(0)]||$,
\begin{eqnarray}
\label{gdefm}
f(t)
\leq \frac{2 N_B S ||A|| ||B|| |2 J S t|^{l/2R} e^{2 J S |t|}}{(l/2R)!}
\\ \nonumber \equiv N_B ||A|| ||B|| g(t,l).
\end{eqnarray}

At this point, from the existence of a $\Delta>0$ and a spectral representation
with all eigenvalues real\cite{tbp}
follows a result similar to Eq.~(\ref{localbnd}):
\begin{eqnarray}
\label{localbndm}
|\langle A(0) B(0) \rangle_c |\leq \frac{1}{2\pi} 
2 N_B ||A|| ||B|| g(c_1 l,l) + \\ \nonumber
2 (1+\frac{1}{\sqrt{2 \pi c_1 l \Delta E}}) ||A|| ||B||
e^{-c_1 \Delta E l/2}.
\end{eqnarray}
Therefore, {\it if} there is a Markov dynamics that gives rise
to the equilibrium probability distribution $p_i^0$ which has a $\Delta>0$,
{\it then} there is an exponential decay of correlation functions in space.
An example is a spin system in the paramagnetic phase with Monte Carlo
spin flip dynamics.  The converse is not necessarily true: a spin system
in the paramagnetic phase with spin exchange dynamics does not have
a $\Delta>0$ but instead has spin correlations which decay with a power law
time.  However, this dynamics
gives rise to the same equilibrium probability distribution
as the spin flip dynamics does, and hence has exponentially decaying
correlations in space.

{\it Acknowledgements---}
This work was supported by DOE contract W-7405-ENG-36.

\begin{figure}[!t]
\begin{center}
\leavevmode
\epsfig{figure=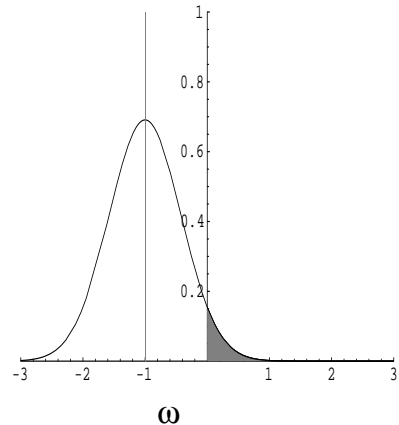,height=8cm,angle=0,scale=.7}
\end{center}
\caption{Illustration of the bound on $|\tilde f^+(0)-f^+(0)|$, as described
in the text.  The vertical line describes a $\delta$-function spike in 
$f(\omega)$.
This produces a Gaussian in $\tilde f(\omega)$.  The integral over $\omega$ of
the Gaussian is the same as the integral of the $\delta$-function; however,
the shaded region of the curve of the Gaussian falls above $\omega=0$, and
hence does not contribute to $f^+(0)$.  This leads to the difference
between $\tilde f^+(0)$ and $f^+(0)$.}
\end{figure}
\begin{figure}[!t]
\begin{center}
\leavevmode
\epsfig{figure=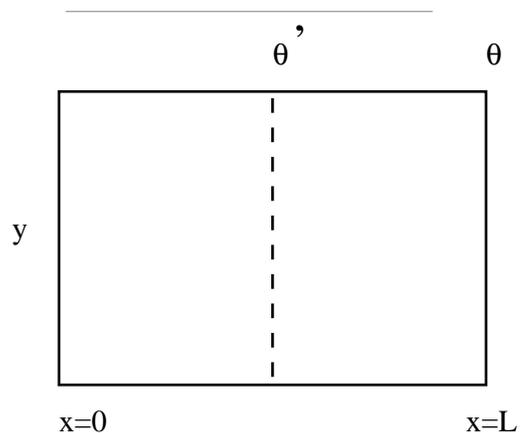,height=8cm,angle=0,scale=.7}
\end{center}
\caption{Plot of the system, showing the $x$-coordinate along the length
axis.  The $x$ is shown ranging from $x=0$ to $x=L$; due to the periodicity
of the system, $x=0$ is identified with $x=L$.  The $y$ coordinate specifies
the position in the directions normal to the length, as well as specifying the
particular site in each unit cell.  The twist angles are noted; the twist 
$\theta$ changes the boundary condition near $x=L$, while
the twist $\theta'$ changes the coupling between sites near $x=L/2$.
}
\end{figure}
\begin{figure}[!t]
\begin{center}
\leavevmode
\epsfig{figure=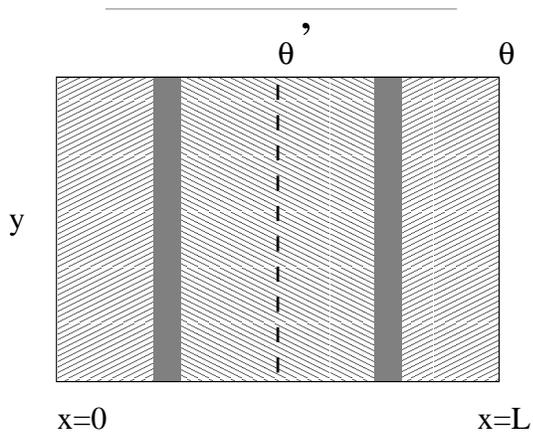,height=8cm,angle=0,scale=.7}
\end{center}
\caption{Plot of the system, showing the twists and coordinates as before.
The halves of the system have been shaded in.  The shading at the left and
right side of the system (diagonal lines going up and right) denotes sites
in half (1), the shading in the middle (diagonal lines going up and left)
denotes sites in half (2).  The solid shading denotes sites in both halves;
the length of the solid region is at least $2R$, so that the Hamiltonian
can be written as a sum of terms, each of which is contained in only
one half.
}
\end{figure}

\begin{thebibliography}{99}
\bibitem{hence}  It follows automatically that Eq.~(\ref{bnd1}) holds for
all $L$ for some $c$ depending on ${\cal H},d,r$.
\bibitem{mdef1} Following standard physics notation, 
whenever we write $1/(-it+\epsilon)$ we
are taking the limit as $\epsilon\rightarrow 0$.  The limit is taken
outside the integral sign.  On the other hand,
the use later of $1/(-it+\tau)$ is done at a fixed, non-zero value of $\tau$.
\bibitem{contour} Eq.~(\ref{intrep}) follows from contour integration.
Another way to derive it is to use a convolution and Fourier transform.
Define the Fourier transform of $f(t)$ to be 
$f(\omega)=\int_{-\infty}^{\infty} {\rm d}t f(t) \exp[-i \omega t]$ 
(see \cite{mdef2}).
Then, $f^+(t)=\frac{1}{2\pi} \int_{-\infty}^{0} {\rm d}\omega f(\omega)
\exp[i \omega t]$.  Combining these gives
$f^+(0)=\frac{1}{2\pi}\int_{-\infty}^{0} {\rm d}\omega \int_{-\infty}^{\infty}
{\rm d}t
f(t) \exp[-i \omega t]$.  Doing the integral over $t$ gives Eq.~(\ref{intrep}).
Note that $\tilde f$ is rapidly decaying as $t\rightarrow\pm\infty$, so
that there are no issues with convergence.
\bibitem{diffable} We note that since we have assumed that the ground
state is unique, with a gap $\Delta E$, then for any finite $L$,
the ground state energy is
analytic in $\theta,\theta'$ in some neighborhood of $\theta+\theta'=0$.
\bibitem{rotg}  To show this, consider rotating all the spins
by angle $\pi$ about
the $x$-axis.  Since the ground state is
assumed to be unique and the Hamiltonian is invariant under
global $SU(2)$ rotations, the ground state is invariant under global
$SU(2)$, so that ground state expectation values must be unchanged under
this rotation.
However, if we take an odd number of derivatives of ${\cal H}_{\theta,\theta'}$
with respect to $\theta,\theta'$, the resulting operator changes sign
under a rotation by $\pi$ about the $x$-axis.  Therefore, the expectation
value of an odd number of derivatives of ${\cal H}_{\theta,\theta'}$ must
vanish.
\bibitem{sl} 
Possible low
energy topological excitations do not prevent the use of Eq.~(\ref{localbnd})
for spin liquids as the matrix elements of
$\partial_\theta {\cal H}_{\theta,\theta'}$ 
between the ground and
topologically excited states are negligible.
\bibitem{tn} The trace norm of an operator $O$, written $|O|$, is
defined to be ${\rm Tr}[\sqrt{O^{\dagger}O}]$.  We will deal with
Hermitian operators, for which
it is equal to the sum of the absolute values of the eigenvalues.
\bibitem{upper} In fact, $|E_1(\theta)|$ need not be differentiable.
To get around this, we use the upper derivative\cite{fgv}, defined as
${\cal D}_t f(t)\equiv{\rm lim \, sup}_{\epsilon\rightarrow 0} 
(f(t+\epsilon)-f(t))/\epsilon$.
\bibitem{h12} We showed before that up to small error $O_1$ is entirely within
half (1).  For larger $c$, we can in fact ensure that $O_1$ includes only
operators within a distance $L/4$ of $x=0$ so that $O_1,O_2$ are separated
by a distance $L/2$ from each other.  Then we can apply the locality bounds
to the given correlation function of $O_1,O_2$.
\bibitem{tbp} It turns out that the requirement that the eigenvalues be
real is necessary.  Although one might have guessed that a
similar exponential decay of correlation functions would also hold
for complex $\lambda$ with real part ${\rm Re}(\lambda)\leq -\Delta$,
there seems to be a counterexample to this statement.  This will
be discussed in a future publication.
\end{thebibliography}
\end{document}